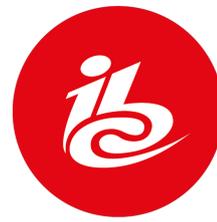

# DIALOG+ IN BROADCASTING: FIRST FIELD TESTS USING DEEP-LEARNING-BASED DIALOGUE ENHANCEMENT

M. Torcoli[1], C. Simon[1], J. Paulus[1,3], D. Straninger[1], A. Riedel[2], V. Koch[2], S. Wirts[2], D. Rieger[1], H. Fuchs[1], C. Uhle[1,3], S. Meltzer[1], A. Murtaza[1]

[1] Fraunhofer Institute for Integrated Circuits IIS, Germany
[2] Westdeutscher Rundfunk Köln (WDR), Germany
[3] International Audio Laboratories Erlangen, Germany

## ABSTRACT

Difficulties in following speech due to loud background sounds are common in broadcasting. Object-based audio, e.g., MPEG-H Audio solves this problem by providing a user-adjustable speech level. While object-based audio is gaining momentum, transitioning to it requires time and effort. Also, lots of content exists, produced and archived outside the object-based workflows. To address this, Fraunhofer IIS has developed a deep-learning solution called Dialog+, capable of enabling speech level personalization also for content with only the final audio tracks available. This paper reports on public field tests evaluating Dialog+, conducted together with Westdeutscher Rundfunk (WDR) and Bayerischer Rundfunk (BR), starting from September 2020. To our knowledge, these are the first large-scale tests of this kind. As part of one of these, a survey with more than 2,000 participants showed that 90% of the people above 60 years old have problems in understanding speech in TV "often" or "very often". Overall, 83% of the participants liked the possibility to switch to Dialog+, including those who do not normally struggle with speech intelligibility. Dialog+ introduces a clear benefit for the audience, filling the gap between object-based broadcasting and traditionally produced material.

## 1. INTRODUCTION

### 1.1 Motivation

Good sound is one of the main challenges in the production of TV shows and movies. It is a complex creative task to produce atmosphere and mood with music and other background sounds in addition to the dialogue. At the same time, the audience wish to fully understand the story and the dialogue in a comfortable way, i.e., without high listening effort.

A portion of the audience served by the German Public Service Broadcasters (PSBs), especially the older population with age-related hearing loss, experiences high listen effort, as demonstrated by the regular complaints addressed to WDR as member of ARD[1]. The most common complaints include loud background noise and music, fast-spoken or overlapping dialogues, and mumbling or unknown dialects. Too loud background sounds were recognized as being a problem already 30 years ago [1].

---

[1] Arbeitsgemeinschaft der öffentlich-rechtlichen Rundfunkanstalten der Bundesrepublik Deutschland (German Association of the Public Service Broadcasters)

PSBs such as WDR are committed to offer accessible services for their audience. Among other efforts to improve accessibility, subtitles are a mean to support, e.g., hard of hearing people. In addition, a new audio track with attenuated background but unchanged overall loudness is currently under evaluation. This paper reports on field tests in which such an additional audio track was automatically generated from the final audio mix by the Dialog+ technology (Sec. 3). Finding the right balance between dialogue and background is the key issue for this extra track, as well for the default mix. This depends on many personal factors, such as hearing acuity, listening environment, language skills, and individual taste [2]. As an alternative, Next Generation Audio (NGA) systems could be a future possibility for WDR and other German PSBs to offer a more personalized audio experience, as introduced in the following Sec. 1.2.

## 1.2 Next Generation Audio (NGA)

NGA systems, such as MPEG-H Audio, introduce new concepts like object-based audio and allow for an increased number of audio channels. This can enhance the user-experience with immersive sound, universal delivery, and advanced interactivity, including accessibility features [3]. Object-based audio uses separate audio elements with attached metadata within one audio stream. One of its most prominent features is dialogue enhancement. In the receiving device, the level of the dialogue can be adjusted automatically based on presets or manually by the user. The minimum and maximum adjustment levels are defined by the broadcaster and transmitted as metadata. Additionally, MPEG-H Audio automatically compensates for loudness changes resulting from user interaction to keep the overall loudness on the same constant level.

MPEG-H Audio has been adopted by the Digital Video Broadcasting Project (DVB) [4] and the Advanced Television Systems Committee (ATSC) standard 3.0 [5], and has been selected by the Telecommunications Technology Association (TTA) in South Korea as the sole audio system for the terrestrial UHDTV broadcasting system [6]. Most recently, it became part of a backward compatible update of the Brazilian Digital Television System (SBTVD) [7]. This update, called TV 2.5, enables broadcasters in Brazil to use one audio stream encoded with AAC as legacy format for existing TV sets, while for newer devices a second audio stream is provided in MPEG-H Audio for offering viewers the most advanced personalization and accessibility features.

Producing content for MPEG-H Audio can be done by using the MPEG-H production tools. These can generate an "MPEG-H Master" file as a bundle of audio and metadata [8]. Existing ADM (Audio Definition Model [9]) productions can also be ingested into MPEG-H production tools and enhanced with more advanced personalization options. Finally, to ensure a consistent user experience, it is essential to offer dialogue enhancement capabilities also for existing content for which the audio objects are not available. This can be enabled by deep-learning-based dialogue separation as done by Dialog+.

Related works are reviewed in the following Sec. 2, while Dialog+ is introduced in Sec. 3. The field tests carried out using Dialog+ are described in Sec. 4. Finally, outlook (Sec. 5) and conclusions (Sec. 6) close this paper.

## 2. RELATED WORKS

### 2.1 Field Tests on Dialogue Level Personalization

One of the very first public trials providing the audience with the possibility of personalizing the dialogue level in broadcast content was carried out by Fraunhofer IIS and the BBC during the 2011 Wimbledon Tennis Championship [10]. Two groups of listener preferences appeared in the trial. The first group preferred to clearly enhance the dialogue, while the second group preferred to slightly reduce the dialogue and to enhance the ambience sounds from the Wimbledon court. Interestingly, the default mix was right in the middle of these two preferences. In other words, the default dialogue level was a good average of the listeners' preferences, but it was preferred by only a small minority once being enabled to choose.

More recently, the BBC and the University of Salford conducted a public trial, where the provided personalization could enhance not only the dialogue, but also narratively important non-speech sounds [11] [12]. A survey accompanied the trial and unveiled that out of the 299 participants 73% found that the personalization feature made the show more enjoyable or easier to understand.

These field tests are based on the assumption that audio objects are available. A complementary line of research aims to enable personalization based on source separation.

### 2.2 Dialogue Separation

Blind source separation methods for extracting the dialogue from broadcast mixture signals have been proposed, e.g., data-driven approaches based on feature extraction and artificial neural networks [13] and more recently deep learning [14].

Other dialogue separation methods are based on the assumption that speech is panned to the center of a 2-channel stereo input signal [15] [16] [17]. An earlier system proposed by some of the authors of this paper was based on combination of classical signal processing algorithms, mostly model-based [18]. This system was able to demonstrate that source separation can offer an increased quality of experience for broadcast content, similarly to the case with audio objects, especially for people older than 65 years [19] [20].

To the best of our knowledge, the field tests described in this paper were the first large-scale public tests using deep learning for enabling audio personalization in broadcasting.

## 3. DIALOG+

Dialog+ is a technology being developed at Fraunhofer IIS. As shown in Figure 1, Dialog+ contains a deep neural network performing dialogue separation. This was trained using supervised regression learning and uses a deep fully-convolutional structure in the short-time Fourier transform domain. The convolutional structure allows for good performance while remaining light to run thanks to the relatively small number of trainable parameters. The system used in the field tests had approximately 370'000 trainable parameters, compared to approximately 10 million in [21] or [22], while still outperforming them in subjective quality [14].

As training data, real-world broadcast content was used, mostly originating from WDR and BR. The content language was mostly German, but a smaller quantity of other languages, such as English and French, were also included. Both female and male speakers were present in the data. The items consist of various content types, from nature documentaries to sports and drama. The raw audio stems were manually cleaned to exclude all examples where non-speech sounds were present in the dialogue stem, or any speech was present in the background stem. This clean-up is very time-consuming, but working with clean, high-quality training material is critical for achieving good results. The clean-up retained only about 15% of the total raw material. The model used in the WDR online test (Sec. 4.1) was trained with approximately 15 hours of stereo audio content at 48 kHz sampling rate. The later field tests (e.g., with BR or the one described in Sec. 4.3) used a model with more data (and minor internal changes), improving the overall quality slightly.

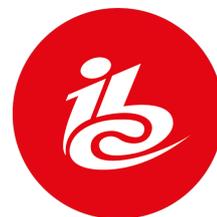

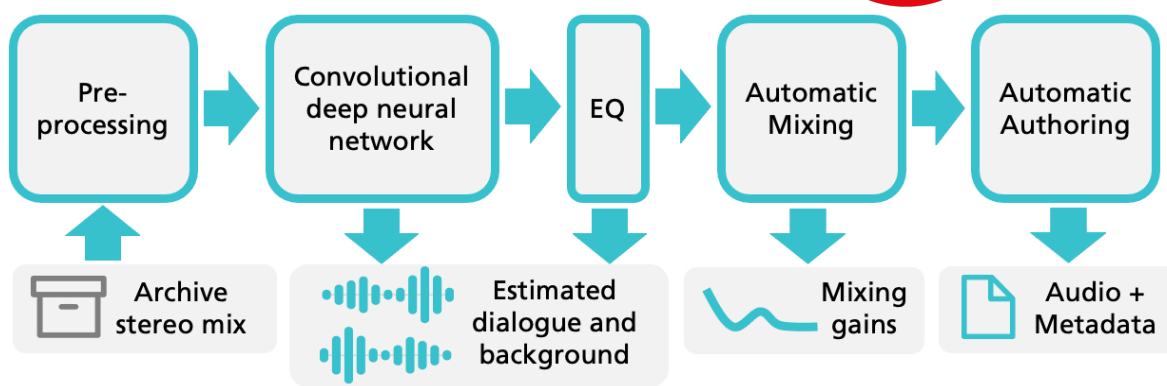

Figure 1 Dialog+ functional blocks along with the products of each of them.

It was observed that the perceptual quality of the separated speech benefits from using a slight constant spectral boost at the main speech frequencies of 1 - 4 kHz in the dialogue component. This boost is compensated in the background component so that the result is identical to the original mix when the components are mixed together without interactivity. This property of preserving the mixture is essential for not compromising the default mix while improving the listening experience in re-mixed conditions.

Dialog+ combines dialogue separation with automatic remixing, where a global and a time-varying background attenuation can be combined. Global background attenuation lowers the relative level of the estimated background component in the same specified amount over the entire signal. This may be beneficial for the users preferring to always lower or even remove the background signal. For others, this might be suboptimal, as attenuating the background while the dialogue is not active does not improve speech intelligibility while potentially damaging mood, atmosphere, and sounds of narrative importance. A solution is to lower the background level only when the dialogue signal is active and only as much as needed for reaching a desired level. The combination of attenuating the background globally by some small amount and more during dialogue activity was used in the reported field tests.

Finally, these outputs can be produced: 1) an ADM file, ready to be used in object-based workflows, and 2) a traditional channel-based audio track with an enhanced dialogue level, used in the reported field tests.

## 4. FIELD TESTS

Between September 2020 and April 2021, three field tests were conducted in cooperation between Fraunhofer IIS and WDR as well as BR, two of the major regional PSBs in Germany. The first field test was carried out in cooperation with WDR, consisting of an online field test followed by a survey, described in Sec. 4.1 and 4.2. The second field test (Sec. 4.3) took place on 13th and 14th December 2020 and it was designed to implement a dialogue enhanced audio channel, called "Klare Sprache" (Clean Speech), to be integrated in the regular DVB services. This scenario enabled the viewers to receive the additional audio channel on a regular TV set.

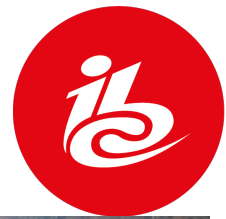

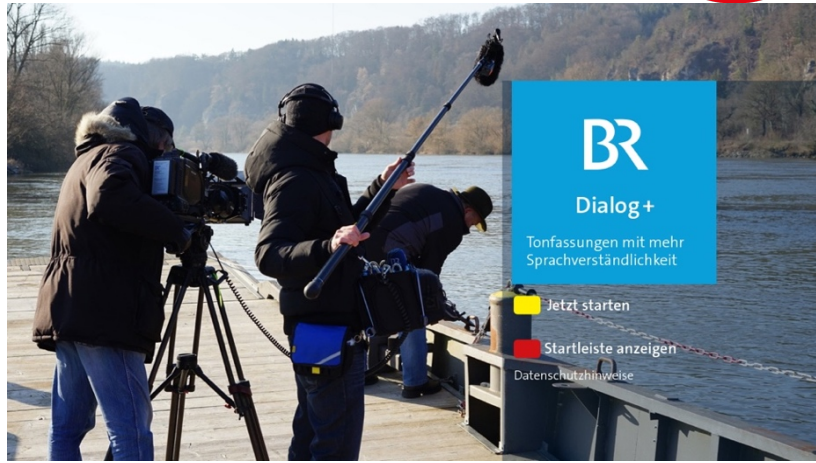

Figure 2 Dialog+ in HbbTV2 during BR field test.
© Bayerischer Rundfunk / Petra Decker, used with permission.

The third field test was conducted in cooperation with BR. It comprised a series of nature documentaries and drama films, which were transmitted regularly over HbbTV2 between December 2020 and April 2021. Here, in addition to selecting the original sound, the user had the opportunity to choose between two different Dialog+ versions with a different degree of background attenuation. The versions were named "Sprache betont" ("speech emphasized") and "Sprache stärker betont" ("speech emphasized more strongly") and were available to be selected over the HbbTV "yellow button" (direct) or via the HbbTV "red button" (HbbTV menu) on the TV remote, as depicted in Figure 2. The HbbTV2 transmission offers the opportunity to broadcast the regular program over DVB, while additional audio versions can be transmitted over the Internet and synchronized to the regular program via the receiver.

For all these field tests, the workflow was semi-automated, i.e., the new audio versions were created automatically by Dialog+, but manually ingested to the transmission chain.

## 4.1 WDR Online Test Setup

WDR and Fraunhofer IIS jointly designed an online test to assess user acceptance for Dialog+. The test took place from 25th September until 30th October 2020. A web page was created and hosted by WDR that integrated the ARD web player version 5.6.0 to switch between audio tracks. This feature is not yet used for ARD regular Video on Demand (VOD) platform, called ARD Mediathek. The player enabled the participants to switch almost seamlessly between the original and the additional audio mix. Figure 3 shows a screenshot of the web page.

Three excerpts from WDR programs with a total length of 3 minutes and 53 seconds were presented: a nature documentary, a program about football, and an episode of a popular police drama series ("Tatort"). These samples represent broadcast content that often cause complaints regarding the dialogue intelligibility. After watching, the participants were asked about their experience and impressions in an online survey. The results from the survey are presented in the following Sec. 4.2.

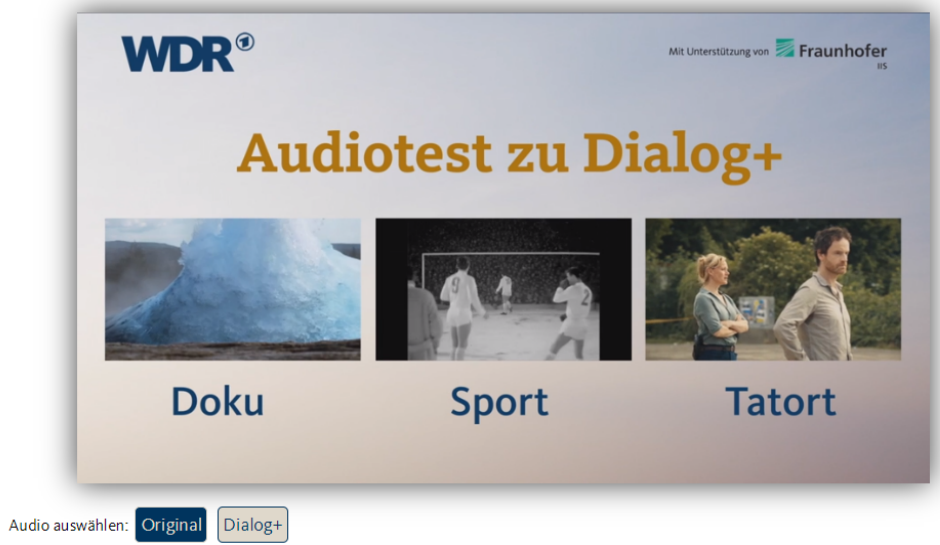

Figure 3 Screenshot of the WDR online test user interface.

## 4.2 WDR Online Test Survey Results

In total, 2126 people participated in the survey, of which 98% answered all questions and 92% did so in less than 5 minutes. Table 1 shows the age distribution of the participants; 79% were between 41 and 80 years old. Figure 4 shows the number of participants over time. Peaks of participation coincided with publication dates of reports about the test in the German news.

Firstly, a better understanding of the importance of the problem was gathered through the question[2]: "*How often do you have difficulties in understanding speech in TV?*"

As Figure 5 depicts, 68% of the participants state that they have problems in understanding speech in TV often or very often. This percentage increases dramatically with age. When considering participants over 60 years old, 90% have problems often or very often. When considering participants over 80, almost the totality (98%) answered often or very often.

| Age groups | Participants | % |
|---|---|---|
| Under 21 | 15 | <1% |
| 21 – 40 | 320 | 15% |
| 41 – 60 | 788 | 37% |
| 61 – 80 | 894 | 42% |
| Above 80 | 64 | 3% |
| Did not say | 45 | 2% |
| Total | 2126 | |

Table 1 - Number of participants per age group. 79% between 41 and 80 years old.

---

[2] The survey was carried out completely in German. This paper presents English translations of the questions and answers in the survey, with the goal of being as close as possible to the original wording.

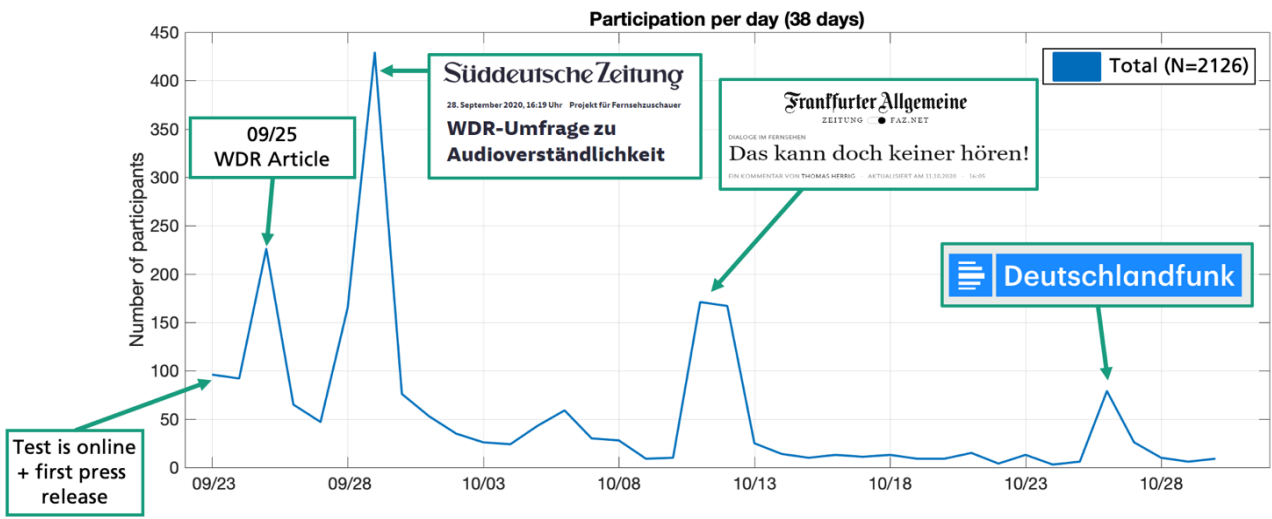

Figure 4 - Participation per day with participation peaks explained by press coverage.

As far as the user experience with the proposed system is concerned, Figure 6 shows the distribution of the answers to the question: "*Did you like the possibility to switch to Dialog+?*" Overall, 83% liked the possibility to switch to Dialog+, including a substantial portion of the people who declared to have never or seldom problems with understanding speech in TV. Both the younger (under 40 years old) and the older population (over 60) responded yes with a great majority (76% and 88% respectively). This clearly indicates that having the possibility to personalize the audio mix is very well received in general, even if this does not address a personal problem or if the original mix is preferred in the end.

When asked "*Which audio version did you prefer?*", 46% responded Dialog+, while 21% preferred the original mix, and the remaining participants had no clear preference, as depicted by Figure 7. Analysing the answers by sub-groups reveals that Dialog+ was preferred in particular by people having difficulties in understanding speech in TV often or very often (60% of them preferred Dialog+), which significantly overlap with the older population (62% of the participants over 60 preferred Dialog+). On the other hand, the original mix was preferred mostly by people not having problems in understanding speech in TV. The original mix was preferred by 48% of participants having intelligibility problems never or seldom. Still, many of them liked the general possibility to switch (Figure 6).

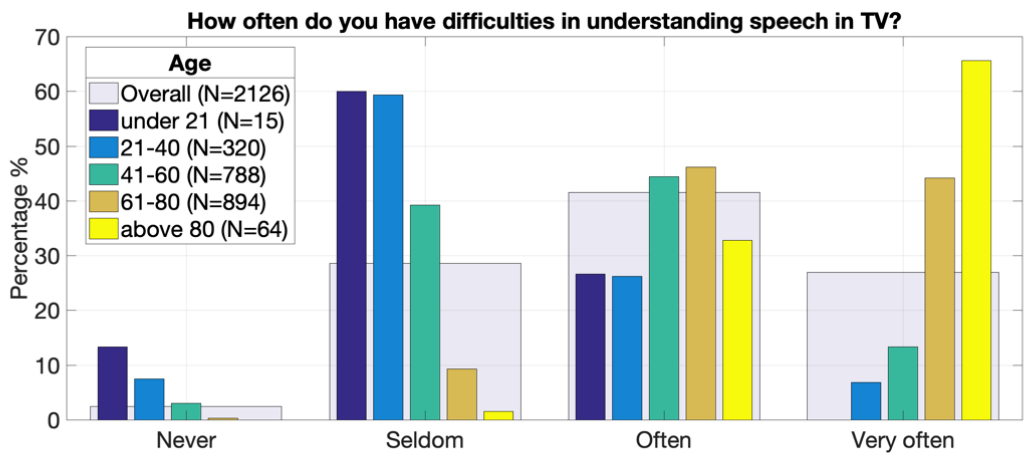

Figure 5 - Problem understanding: 68% of the participants have problems in understanding speech in TV often or very often (i.e., summing up the two right-most grey bars). The percentage increases with age. Over 60 years old, 90% have problems often or very often.

Participants were also asked to motivate their preference choice through multiple predefined replies and an optional open text field. The most selected motivations for preferring Dialog+ were: *Dialogue is easier to understand* (selected by 80% of the participants preferring Dialog+) and *Better balance between dialogue and background noises and music* (78%). The most selected motivations for preferring the original mix were: *Original mix sounds more natural* (selected by 76% of the participants preferring the original mix) and *Better sound quality* (62%). As far as the participants without a clear preference are concerned, the main motivation was that the audio versions were too similar (42%). This motivation was particularly popular among the participants using integrated laptop speakers (51%), while it was less significant for the participants using TV loudspeakers (28%) or headphones (29%).

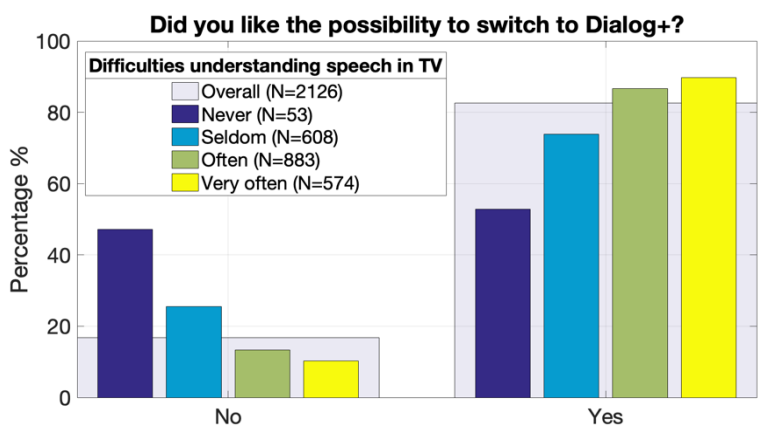

Figure 6 - Overall, 83% liked the possibility to switch to Dialog+, including those who do not normally struggle with speech intelligibility.

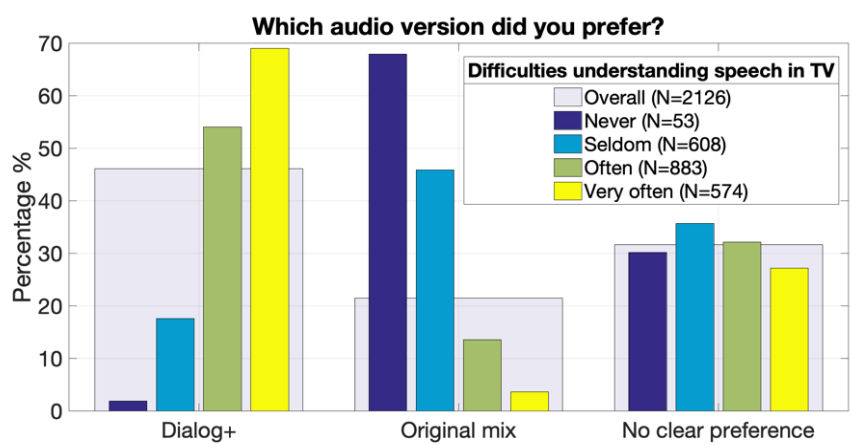

Figure 7 - Overall, 46% preferred Dialog+, while 21% preferred the original mix and the remaining participants had no clear preference.

These motivations behind the choice between Dialog+, original mix, or no preference were also confirmed by analysing the hundreds of messages given in the open text fields. The analysis of these texts was done as follows:

1. Key words were manually identified in the texts (e.g., "Improvement", "Enhancement", "Speech Intelligibility", "Unnatural sounding").

2. Key words referring to the same concept were aggregated (e.g., "Improvement" and "Enhancement" were aggregated into "Improvement").

3. The frequency of the main concepts was computed and converted into a font size.

4. The word clouds in Figure 8 and Figure 9 were created. In these clouds, the bigger the font of a concept is, the more often that concept was encountered in the textual feedback messages from the participants.

Especially the word cloud *Contra Dialog+* (Figure 8) gives us valuable information for steering the further development of Dialog+ and its core dialogue separation. While people struggling with speech intelligibility in TV are already convinced by the system, a better separation quality seems to be needed to additionally convince the people who do not normally struggle with intelligibility. In the meantime, personalization of the audio mix is once again shown to be the only way to meet the taste and needs of everyone in the audience.

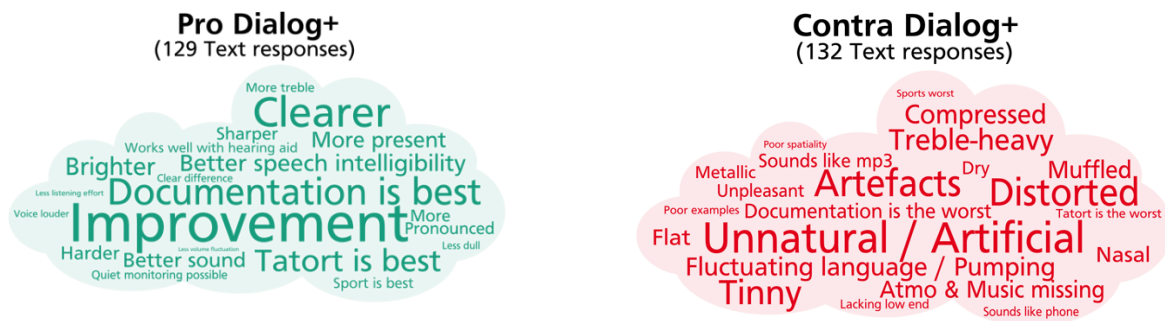

Figure 8 - Word clouds visualizing participants' opinions expressed via the open text responses. The cloud on the left shows the opinions with a positive tendency towards the Dialog+ version, while the right cloud shows the opinions with a negative one.

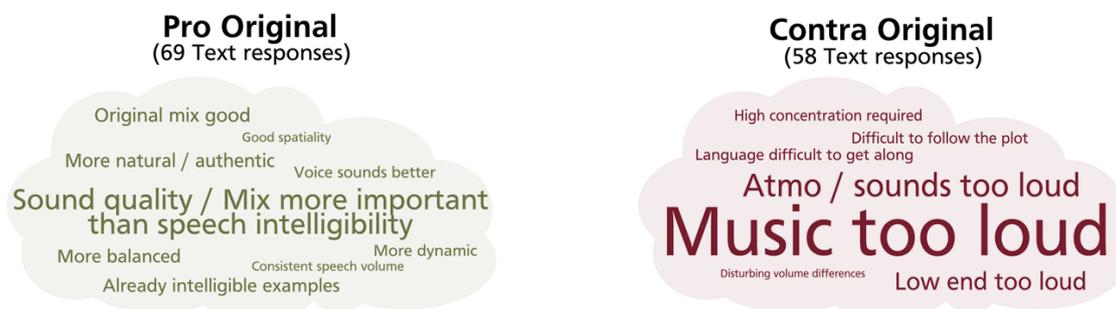

Figure 9 - Word clouds visualizing participants' opinions expressed via the open text responses. The cloud on the left shows the opinions with a positive tendency towards the original mix, while the right cloud shows the opinions with a negative one.

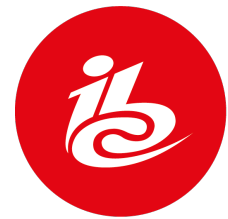

## 4.3 WDR "Klare Sprache" over DVB and streaming

Following up on the positive feedback received in the online test, WDR and Fraunhofer IIS organized a second field test with regular DVB broadcast. This provided the audience of the TV channel "WDR HD Köln" with an additional audio track over DVB-S and DVB-C.

A short hint was displayed overlaying the video referring to a teletext page with more information about the additional audio channel for "Klare Sprache" (Clean Speech) which could be selected using the audio menu of the TV set, as shown in Figure 10. A total of nine fictional films and documentaries was broadcast in this way. A selection of these programs was also made available for streaming in the ARD VOD service.

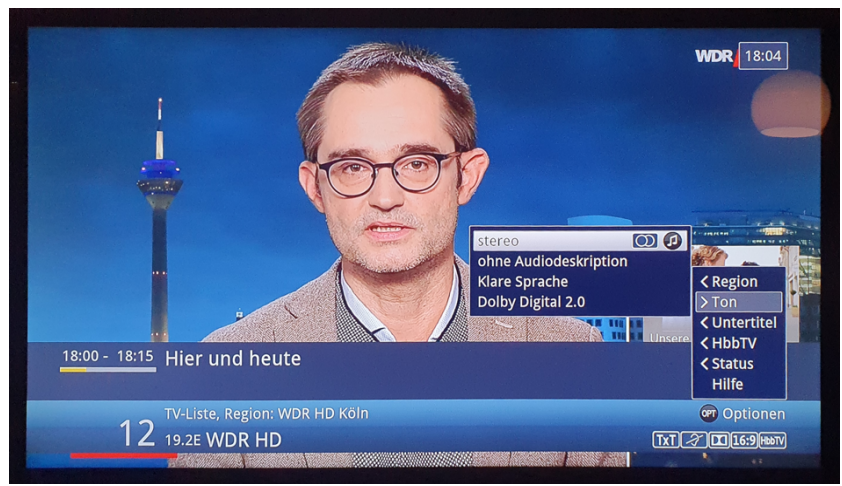

Figure 10 - WDR Klare Sprache (Clean Speech) over DVB.

© Alfred Riedel, WDR, used with permission.

## 5. OUTLOOK

For broadcasters, it is important to offer to the audience more accessible audio through a process that is as much automated as possible. This applies to both the 24/7 live production process and increasingly to the on-demand services. Especially live production is considered to be a remaining challenge. In any case the whole audience should be provided with this additional audio service called "Klare Sprache" (Clean Speech). Ambition of WDR is to enable an easy and intuitive access to the new audio signal via the audio menu of the receiving device. Dialog+ proved to be able to deliver this and to be well received by the audience.

One participant of the field tests wrote: "For me, 82 years old, Dialog+ would be a substantial improvement of the speech intelligibility! For years now I can hardly watch any film anymore because of my hearing loss. I hope that Dialog+ will establish itself, millions of elderly people would be very thankful!" An answer to this participant comes from a WDR press release [23]: "Also considering the accessibility requirements of our service, there is enough motivation to further develop such a service."

Dialog+ and its implementation in production tools are novel and parts of them are still under development. Opportunities for improvement are plenty, e.g., advancements on the signal processing side as well as on the production side regarding tools and workflows. Additional features are under experimentation, e.g., automatic quality control algorithms [24] [25].

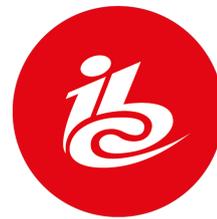

As one of the next steps for the further development of the technology, Fraunhofer IIS has started an evaluation program for Dialog+ allowing international broadcasters and producers to test the Dialog+ implementation. The participants in the evaluation program can give feedback about the audio quality, envisioned workflows and system integration. The results will be incorporated into the further development of Dialog+ and accompanying tools.

## 6. CONCLUSIONS

This paper reported on public field tests with Dialog+ via DVB, HbbTV, and online streaming. Dialog+ is a solution for speech level personalization that uses deep learning to enable the adjustment of the dialogue level when only the final audio mix is available.

As part of one of these field tests, a survey with more than 2,000 participants showed that 90% of the people above 60 years have problems in understanding speech in TV often or very often. Overall, 83% of the participants liked the possibility to switch to Dialog+, including those who do not normally struggle with speech intelligibility. Dialog+ proved to bring a clear benefit for the audience, filling the gap between object-based broadcasting and traditionally produced material.

## ACKNOWLEDGEMENTS

The authors would like to warmly thank Kathrin Weinzierl and her colleagues at BR for their support and cooperation.

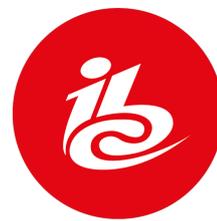